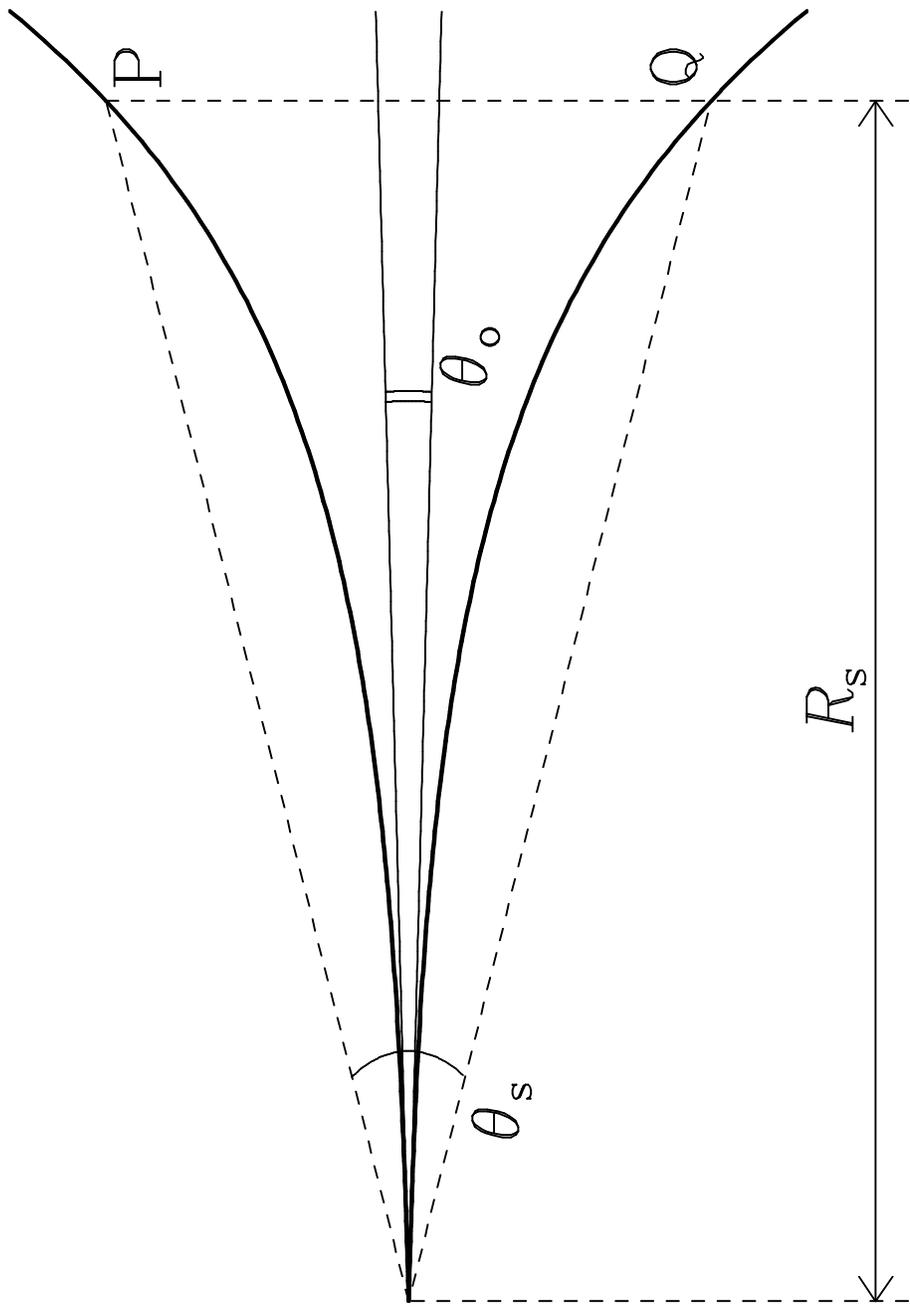

# EXPONENTIAL GROWTH OF DISTANCE BETWEEN NEARBY RAYS DUE TO MULTIPLE GRAVITATIONAL SCATTERINGS

Toshiyuki Fukushige and Junichiro Makino[†]

Department of Earth Science and Astronomy,
[†]Department of Information Science and Graphics,
College of Arts and Sciences, University of Tokyo,
Komaba 3-8-1, Meguro-ku, Tokyo 153, Japan
E-mail:fukushig@chianti.c.u-tokyo.ac.jp

**Abstract**

We give an estimate of the relative error in the angular measurement of observations for high redshift objects induced by gravitational scatterings (lensing). Gunn (1967) concluded that the gravitational scatterings by galaxies induce the relative error of a few percent in the observations for objects at $z = 1$. This estimate has been considered as a fundamental limitation of accuracy of the angular measurements in the observational cosmology. In multiple gravitational scatterings, bending angle of single ray grows through the random work process. Gunn (1967) assumed that the difference of nearby rays also grows through the random walk process. However, distance between nearby photons grows exponentially because the two rays suffer coherent scatterings by the same scattering object. This exponential growth continues as long as the scattering is coherent. In the case of scattering by individual galaxies, the exponential growth continues until the angular distance reaches an arcminute or so. The relative error of the angular measurements under an arcminute due to the exponential growth is $\sim 30\%$ at $z = 1$ and exceeds $100\%$ at $z = 3$, in the case that the density parameter of galaxies is 0.2. The effects of clusters of galaxies or superclusters are more difficult to estimate accurately, but might be significant. In the case of supercluster the angular measurements up to a few degrees could be affected.
Subject headings: cosmology: observations, gravitational lensing

## 1. Introduction and Summary

A photon is gravitationally scattered by galaxies during its travel from its source to us. By this effect of the gravitational scattering (lensing), an image of high redshift object can be distorted to a double-image and an arc, or can be attenuated (Blandford and Narayan 1992, and references therein). In this letter, we calculate the relative deflection of two photons, and investigate change in the observed angle of the high redshift object induced by the gravitational scatterings. The change in the angle puts an limitation on the angular measurement of the observation of the high redshift objects, and gives the change of their size and apparent magnitude approximately. We found that the change due to the gravitational scattering is larger than the estimate of the previous researchers (e.g. Gunn 1967).

Figure 1 shows the basic idea. We observe two photons with the relative angle of $\theta_o$. They are actually originated from points $P$ and $Q$ at $z = z_s$, and the real angular distance between them is $\theta_s$. The trajectories of the photons are deflected by the gravitational scatterings, resulting in $\theta_o \neq \theta_s$. In order to evaluate the difference between $\theta_o$ and $\theta_s$, we consider the following situation: The observer "sees" two photons with angle $\theta_o$ at present. We calculate the orbit backward in time and determine from where they come. We define a "demagnification" factor, $\alpha(z)$, between two angles as $\alpha(z) \equiv \theta_s(z)/\theta_o$. For two points with angular distance $\theta_s$, the observed angular distance becomes $\theta_o$.

Gunn (1967) calculated analytically the relative deflection between two rays due to the gravitational scattering, and concluded that the relative change of observed angle induced by the gravitational scattering is a few percent for the object at redshift $z = 1$. This estimate has been regarded as a fundamental limitation of accuracy of the angular measurements in the observational cosmology. Kantowski (1969), Dyer and Roeder (1974) and Dyer and Oattes (1988) investigated distortion of distant objects using the Swiss-cheese model. Blandford and Jaroszynski (1981) modified the calculation of Gunn (1967) using two-point correlation function of galaxies that had newly measured, and Watanabe and Tomita (1990) calculated the relative deflection among rays using the result of the numerical simulation of light propagation by Tomita and Watanabe (1989). They obtained results similar to Gunn (1967)'s result.

In the previous theoretical estimates, it was assumed that the two neighboring rays was described by simple behaviour of the random walk. However, the effect of multiple scatterings on the neighboring rays can not be described by the random walk. In fact, the distance between two rays grows exponentially due to the multiple gravitational scatterings. In this letter, we estimate the change in the angle due to the exponential growth.

A basic explanation of the exponential growth of distance is as follows. More detailed discussions have been given by Goodman, Heggie and Hut (1993). When the distance between two photons is small, the photons are scattered coherently by single scattering object. In this coherent scattering, the distance increases because the photon closer to the scattering object is deflected with a larger angle. When they are scattered in the next time, the distance increases again. The scattering angle $\theta$ is inversely proportional to the impact parameter. It is expressed as $\theta \simeq 4GM/(bc)^2$, where $G$ is the gravitational constant, $M$ is the mass of the scattering object, $b$ is the impact parameter, and $c$ is the light velocity.

The difference in the scattering angle is given by

$$\Delta\theta = w\frac{d\theta}{db} \simeq \frac{4GM}{b^2 c^2}w, \quad (1)$$

where $w$ is the distance between the two photons. Equation (1) implies $\Delta\theta \propto w$. On average, the distance between nearby photons grows exponentially due to the multiple scatterings, as long as the scatterings are coherent. Figure 2 illustrates this exponential growth of the distance between two nearby photons.

In the community of $N$-body simulation, the exponential instability of initial condition has been well known (Miller 1964; Lecar 1968; Sakagami and Gouda 1990; Suto 1990; Kandrup and Smith 1991; Kandrup, Smith and Willmes 1992; Quinlan and Tremaine 1992; Huang, Dubinski and Carlberg 1993; Goodman, Heggie and Hut 1993). According to Goodman, Heggie and Hut (1993), the e-folding time of this exponential instability is of the order of 0.1 crossing times, and the exponential growth would saturate when the distance between two orbits becomes $RN^{-1/2}$, where $R$ is the characteristic radius of the system and $N$ is the number of particles in the system. On the view of light propagation, the result of Tomita and Watanabe (1989) showed the exponential growth of the initially small angle in the numerical simulation including cosmological expansion. Fukushige, Makino and Ebisuzaki (1994) and Fukushige et al.(1994) discussed the effect of this exponential instability on the observed temperature anisotropy of the cosmic background radiation.

This exponential growth stops when the scatterings become incoherent. The condition for the exponential growth is

$$w < w_{\rm cr} \sim \left(\frac{GM}{c^2 n}\right)^{\frac{1}{4}}, \quad (2)$$

where $w_{\rm cr}$ is the critical distance that the exponential growth occurs and $n$ is the number density of the scattering objects. If we consider the case of $\Omega_{\rm s} = 1$, where $\Omega_{\rm s}$ is the density parameter of the scattering objects, the critical distance at $z = 0$ can be written as $w_{\rm cr} \sim R_{\rm H} N^{-\frac{1}{2}}$ where $R_{\rm H}$ is the distance to the horizon and $N$ is the number of the scattering objects. After the distance becomes larger than this critical distance, the distance grows diffusively. In this region of the diffusive growth, we should use Gunn's (1967) formula.

In this letter we estimate the change in the angle between two rays due to the gravitational scattering. Our main conclusions are summarized as follows. The demagnification factor defined as the ratio between the true and observed angular size of an object at redshift $z$ is expressed as

$$\alpha(z) \simeq \frac{1}{1 + 1.28\sqrt{\Omega_{\rm s}}} \left[\frac{(1+z)^{1.92\sqrt{\Omega_{\rm s}}} - (1+z)^{-3/2}}{1 - (1+z)^{-3/2}}\right]. \quad (3)$$

For individual galaxies, $\Omega_s = 0.2$ would be a reasonable estimate. In this case, $\alpha \sim 2$ for redshift $z = 3$, and $\alpha \sim 3$ for $z = 6$. This growth continues until the distance between rays becomes similar to the critical distance $w_{cr}$. For the case of galaxies, $w_{cr} \sim 1$ arcminute. This means that the error is as large as 100% in the angular measurement with angular size smaller than an arcminute, which corresponds to a size of $\sim 0.5 h^{-1}$Mpc at $z = 1$, where $h^{-1}$ is the Hubble constant in the unit of 100km/(s·Mpc).

For this exponential growth to occur, the scattering objects must be more compact than the critical size. Galaxies are compact enough to be effective as the scattering object. The critical size for the scattering objects is the critical distance $w_{cr}$. If the half-mass radius of the scattering object, $r_h$, is smaller than the critical distance $w_{cr}$, the exponential growth occurs. This is because main contribution of the exponential growth is due to the scatterings having impact parameters of the order of $w_{cr}$ (Goodman, Heggie and Hut 1993). Numerical calculations confirmed the saturation of the growth factor at larger $r_h$ than the critical distance of $w_{cr}$ (Fukushige, Makino and Ebisuzaki 1994; Fukushige et al. 1994). The critical distance $w_{cr}$ of galaxies is estimated as $\sim 0.5$Mpc in the case of mean separation of galaxies: $d_{sep} = 3$Mpc and $\Omega_s = 0.2$. Therefore, the exponential growth due to the gravitational scattering by galaxies occurs, unless a dark halo of galaxies spreads up to a size of $\sim 0.5$Mpc.

None of the previous studies took into account this exponential growth of the distance between rays. They estimated the effect of the multiple scatterings by a superposition of the single scatterings. For example, Gunn (1967) calculated the observed relative deflection, **n**, between two rays, using the expression

$$\mathbf{n} = 4GM \int \left[ \frac{\mathbf{D} + \mathbf{d}}{(D+d)^2 + \varepsilon^2} - \frac{\mathbf{D}}{D^2 + \varepsilon^2} \right] \frac{l_{ex}}{l_{e0}} dN(x), \tag{4}$$

in equation (12) of his paper, where $D$ is the impact parameter, $d$ is the difference of the impact parameter, $G$ is the gravitational constant, $\varepsilon$ is the size of the scattering objects, $dN(x)$ is the number of galaxies at **x**, and $l_{ex}$ is the distance from the emitter to **x**. In equation (4) the total effect of the multiple scatterings is calculated as simple summation of many scatterings with the distance between two rays, $d$, unchanged. Therefore, the possibility of the exponential growth is precluded in equation (4).

## 2. Change of Angle due to the Gravitational Scattering

In this section, we estimate the demagnification factor $\alpha(z)$. Here, we assume that the universe is flat ($\Omega = 1$) for simplicity. The angle $\theta_s(z)$ is given by $\theta_s(z) = w(z)/R_s$, where $w(z)$ is the distance between photons at redshift $z$, and $R_s$ is the distance from $z = 0$ to redshift $z$. The evolution of the distance $w$ is approximately described by the equation:

$$\frac{dw(z)}{dt} \simeq -c\theta_{\rm o} - \frac{w(z)}{t_{\rm e}(z)}, \qquad w(z=0) = 0, \tag{5}$$

where $t_{\rm e}$ is the e-folding time of the exponential growth of the distance between two rays. The e-folding time $t_{\rm e}$ is given by $t_{\rm e} = \tau/\sqrt{G\rho}$, where $\rho$ is the mass density of the scattering objects, and $\tau = 0.18$, according to the numerical simulations of Fukushige, Makino and Ebisuzaki (1994). In an expanding universe, the e-folding time is expressed as

$$t_{\rm e}(z) = \tau \left(\frac{8\pi}{3H_0^2 \Omega_{\rm s}}\right)^{1/2} (1+z)^{-3/2}, \tag{6}$$

where $H_0$ is the Hubble constant. By solving equation (5) with (6), we obtain $\theta_{\rm s}(z)$ as

$$\theta_{\rm s}(z) \simeq \frac{\theta_{\rm o}}{1+\eta} \frac{(1+z)^{3\eta/2} - (1+z)^{-3/2}}{1 - (1+z)^{-3/2}}, \tag{7}$$

where $\eta = \sqrt{\Omega_{\rm s}}(6\pi)^{-1/2}\tau^{-1}$. Figure 3 shows the demanification factor $\alpha(z)$ as a function of redshift $z$ for $\Omega_{\rm s} = 0.2$ (the solid curve) and $\Omega_{\rm s} = 1$ (the dashed curve). In the case of $\Omega_s = 0.2$, the angle at $z = 0$ increases by a factor 2 at $z = 2$ and by a factor 3 at $z = 5$. If total mass of the universe is in galaxies, i.e., $\Omega_s = 1$, $\alpha \simeq 5$ at $z = 2$ and $\alpha \simeq 15$ at $z = 5$.

Figure 4 shows $\theta_{\rm s}(z)$ as a function of $z$ for different values of $\theta_{\rm o}$ for the case of $\Omega_{\rm s} = 0.2$ and $d_{\rm sep} = 3h^{-1}{\rm Mpc}$. The thick curve shows the angular size, $\theta_{\rm cr}(z)$, of the object with the size equal to the critical distance $w_{\rm cr}$. The angle $\theta_{\rm cr}(z)$ is given by

$$\theta_{\rm cr}(z) \simeq \frac{w_{\rm cr}}{R_{\rm s}(z)} \simeq 20" \cdot \Omega_{\rm s}^{-1/4} \cdot \frac{(1+z)^{-3/4}}{1 - (1+z)^{-3/2}}. \tag{8}$$

If $\theta < \theta_{\rm cr}$, the change of $\theta$ is given by equation (7); If $\theta > \theta_{\rm cr}$, the angle $\theta$ changes by the random walk.

Gunn's (1967) estimate of the error in the angular measurements, given in his equation (41a), is

$$\left[\frac{(\theta_{\rm s} - \theta_{\rm o})^2}{\theta_{\rm o}^2}\right]^{\frac{1}{2}} \simeq \left(\frac{8C}{30}\right)^{\frac{1}{2}} [1 - (1+z)^{-1/2}]^{3/2}, \tag{9}$$

where $C = 2.9 \times 10^{-2}$. Figure 5 shows the errors in the angular measurement as a function of redshift $z$. The dashed curve is the error estimated by Gunn (1967) for $\Omega_{\rm s} = 1$. Our estimate is larger than that of Gunn (1967) by more than two orders of magnitude.

## 3. Discussion

We estimated the effect of the gravitational scattering using the factor $\alpha$ between the true and observed angular size. Using this factor $\alpha$, we can approximately estimate increase of the apparent magnitude of the high redshift sources due to the gravitational scattering. Since the surface brightness of the source doesn't change by the gravitational scattering, the apparent magnitude is determined by the angular size of the source. If we assume that the size of a typical object shrinks by a factor $1/\alpha(z)^2$, the change $\Delta m$ in the apparent magnitude $m$ is

$$\Delta m = m_0 - m = \frac{5}{2}\log_{10}\alpha(z)^2$$
$$= \begin{cases} 5\log_{10}\left[\dfrac{(1+z)^{0.86} - (1+z)^{-3/2}}{1 - (1+z)^{-3/2}}\right] - 0.98, & (\Omega_{\rm s} = 0.2) \\ 5\log_{10}\left[\dfrac{(1+z)^{1.92} - (1+z)^{-3/2}}{1 - (1+z)^{-3/2}}\right] - 1.79, & (\Omega_{\rm s} = 1) \end{cases} \quad (10)$$

where $m_0$ is the apparent magnitude for the case without the gravitational scattering. The apparent magnitude of a source at $z = 5$ increases by 2.5 for $\Omega_{\rm s} = 0.2$ and by 5.8 for $\Omega_{\rm s} = 1$. This effect certainly affects the magnitude-redshift relation. Also, this effect may be related to cut-off in the distribution of quasars at high $z$.

Finally, we estimate the effect of clusters of galaxies and superclusters. Both cluster of galaxies and superclusters are compact enough to work effectively as the scattering object of the exponential growth. APM survey for cluster of galaxies (Dalton et al 1992) showed that the mean separation of clusters is $34h^{-1}$Mpc. The thickness of the grate wall in CfA survey (Gellar and Huchra 1989) is $5h^{-1}$Mpc, while the size of a void is $\sim 100h^{-1}$Mpc. However, the density parameter of cluster of galaxies and supercluster has a large uncertainty. Also, the formation time of these object, or their very existence, is still controversial. Because of the larger average separation, the critical angle $\theta_{\rm cr}$ is larger for clusters of galaxies or superclusters than for galaxies. For superclusters, $\theta_{\rm cr}$ is as large as a few degrees.

We thank Toshikazu Ebisuzaki and Yoko Funato for helpful discussion. This research is partially supported by the Grand-in-Aid for Specially Promoted Research (04102002) of The Ministry of Education, Science, and Culture. This study was carried out while TF was a fellow of the Japan Society for Promotion of Science for Japanese Junior Scientists.

**Figure legend**

**Figure 1** The change in angle between two rays emitted from the observer. The angles $\theta_s$ and $\theta_o$ are the true and observed angular size, repectively.

**Figure 2** The exponential growth of distance between nearby rays due to multiple gravitational scatterings.

**Figure 3** The demanification factor $\alpha(z)$ as a function of redshift $z$. The solid curve is for the case of $\Omega_s = 0.2$, where $\Omega_s$ is the density parameter of galaxies, and the dotted curve is for $\Omega_s = 1$.

**Figure 4** The true angular size $\theta_s(z)$ for different observed angles at $z = 0$, plotted as a function of $z$. The density parameter is $\Omega_s = 0.2$. The tick curve indicates the true angular size, $\theta_{cr}(z)$, of the object with the size $w_{cr}$ at redshift $z$. If $\theta < \theta_{cr}$, the change of $\theta$ is described by the exponential growth.

**Figure 5** The error ($[(\theta_s(z)-\theta_o)^2/\theta_o^2]^{1/2}$) in angular measurement on observation for high redshift objects. The solid and dotted curves show our estimate for the case of $\Omega_s = 0.2$ and $\Omega_s = 1$, respectively. The dashed curve shows the error estimated by Gunn (1967) for $\Omega_s = 1$.

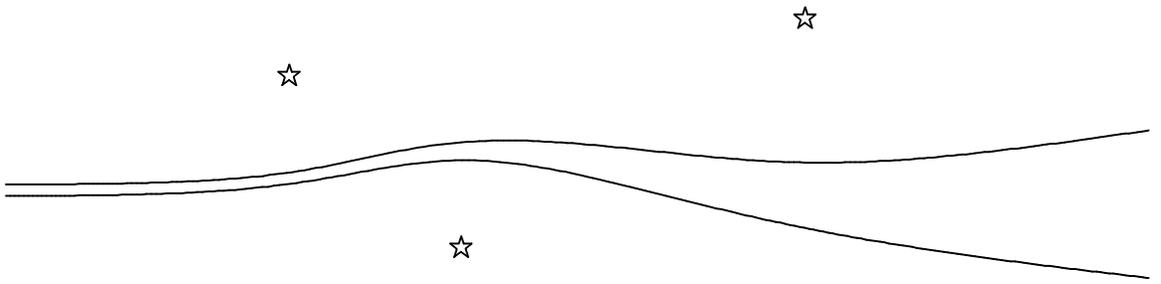

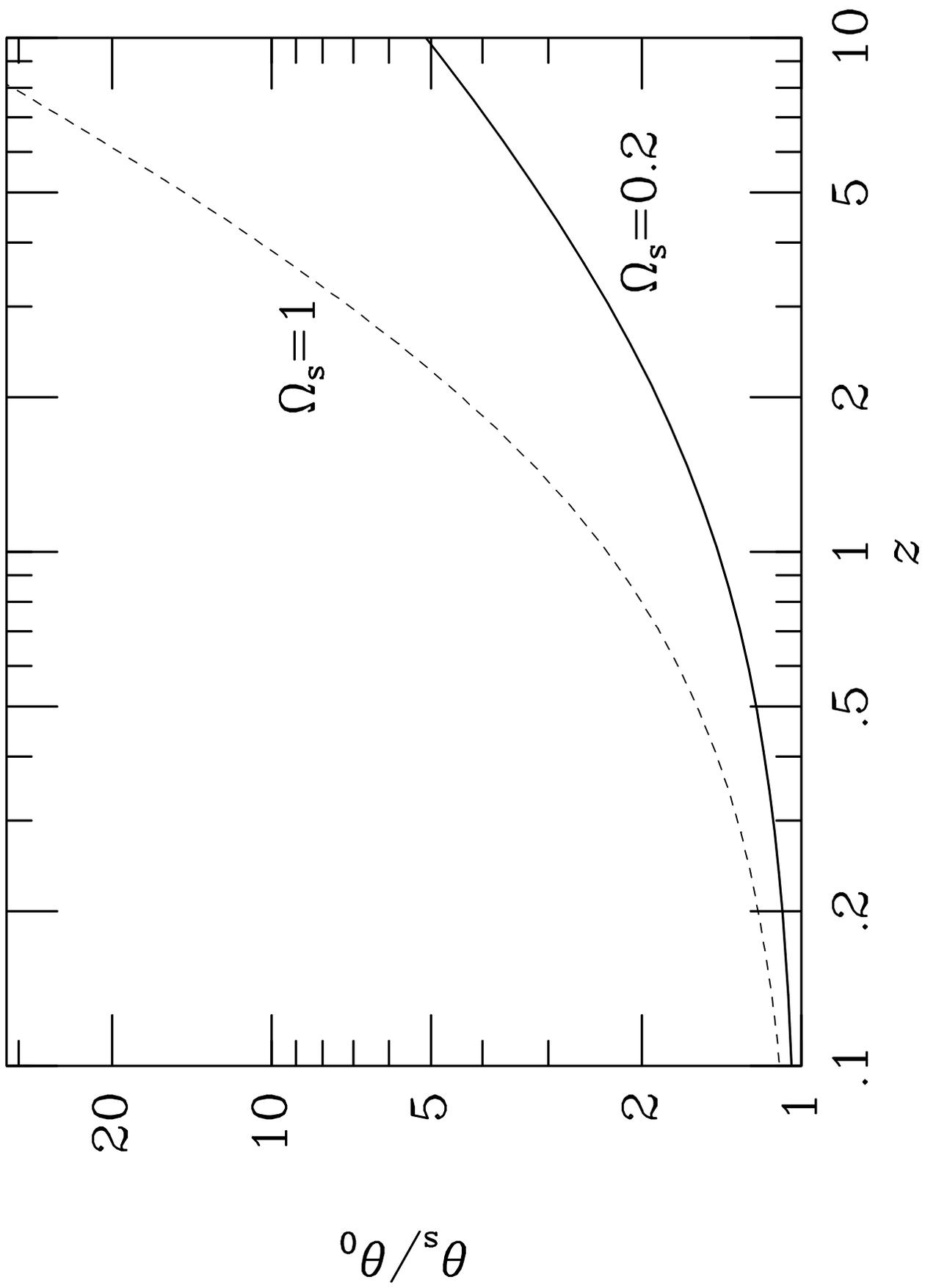

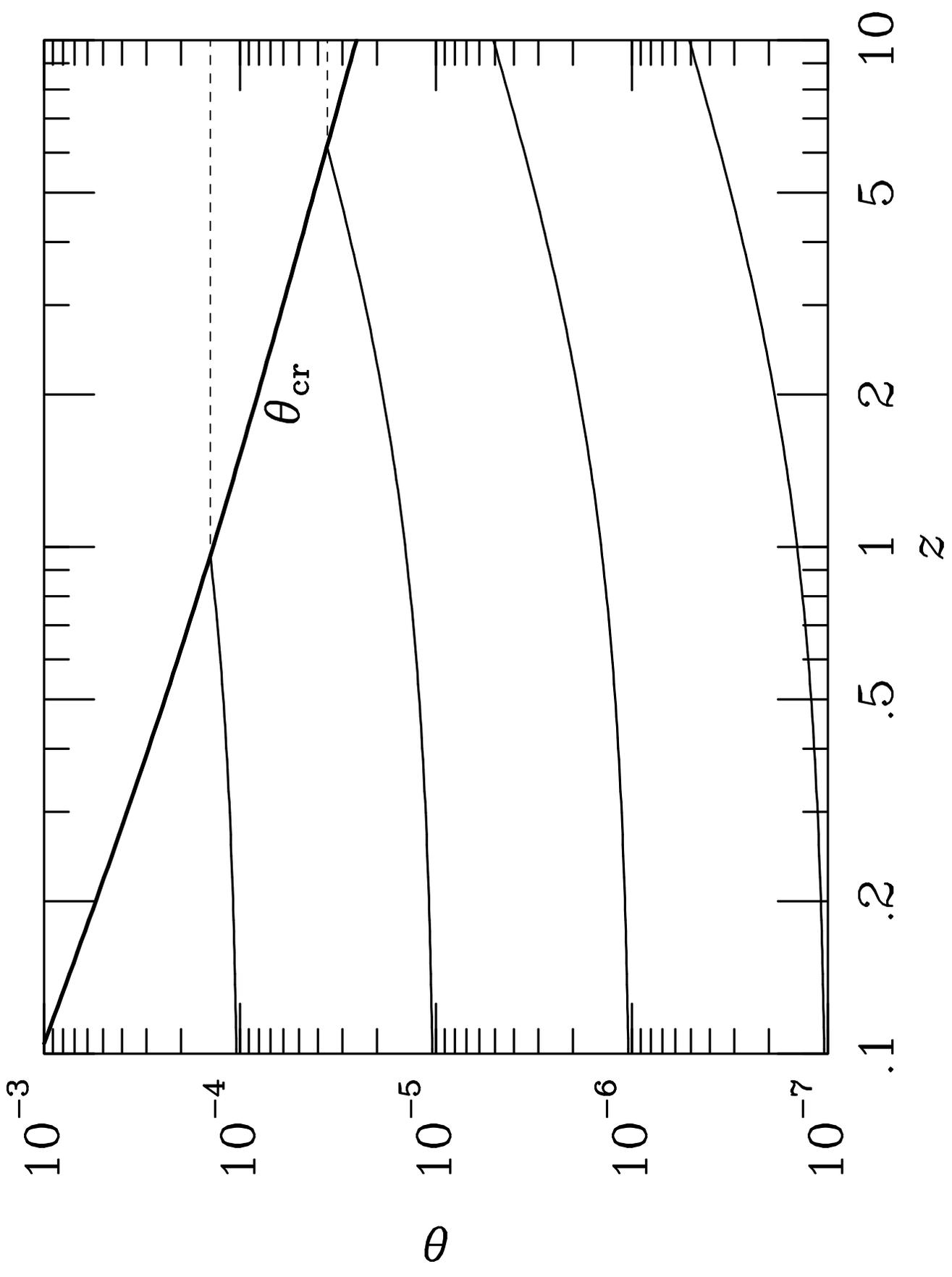

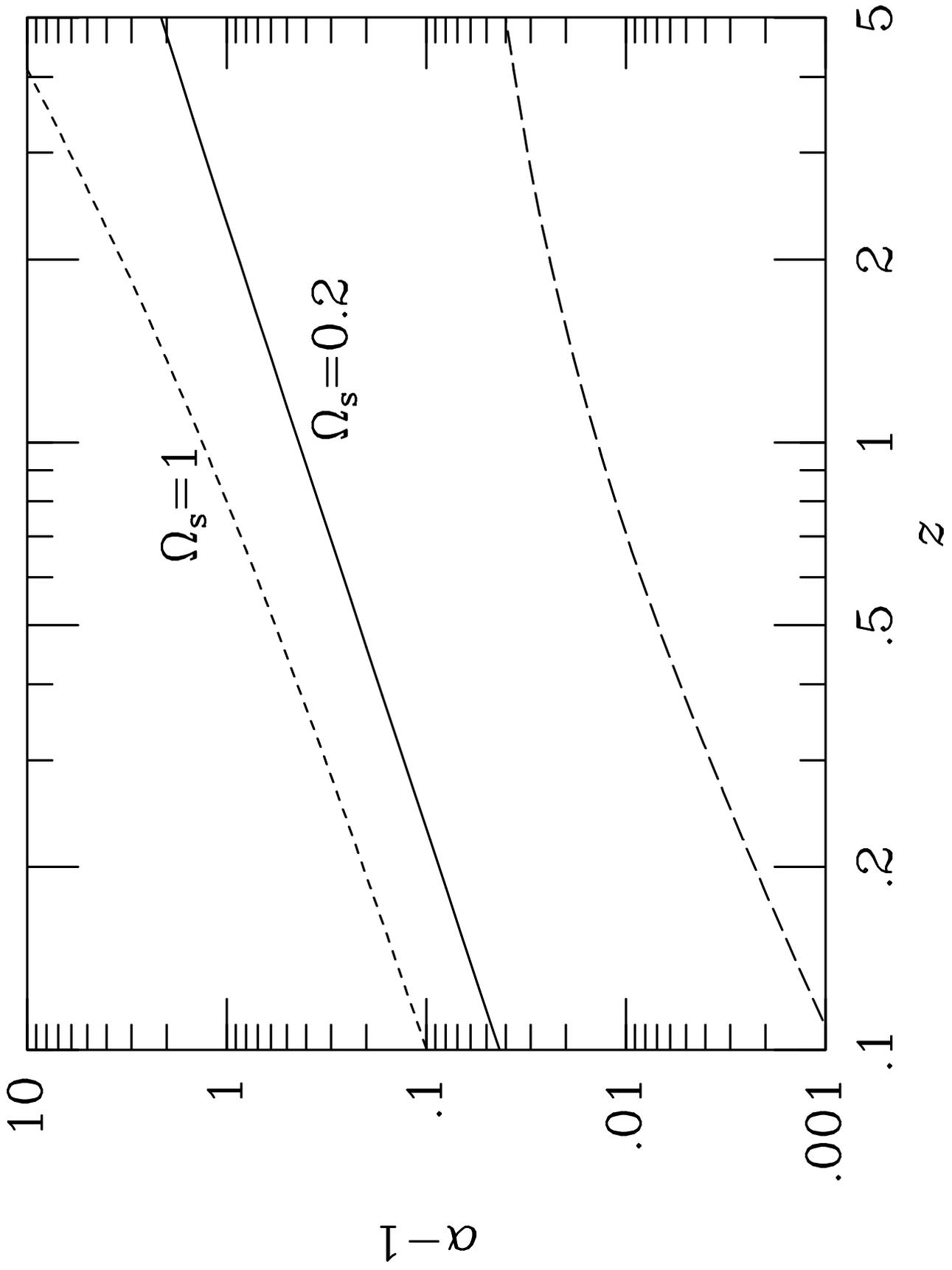